\documentclass[12pt]{iopart}
\usepackage{graphicx}
\usepackage[dvips]{color}

\newcommand{\eff}{\mbox{\scriptsize eff}}

\begin{document}

\title[Dynamic heterogeneities, fragility, and slow dynamics in random media]
{Slow dynamics, dynamic heterogeneities, and fragility of supercooled liquids confined in random media}

\author{Kang Kim$^{1, 2}$, Kunimasa Miyazaki$^3$, and Shinji Saito$^{1, 2}$}
\address{$^1$Institute for Molecular Science, Okazaki 444-8585, Japan, \\
$^2$The Graduate University for Advanced Studies, Okazaki 444-8585, Japan, \\
$^3$Institute of Physics, University of Tsukuba, Tsukuba 305-8574, Japan}
\ead{kin@ims.ac.jp}

\begin{abstract}
Using molecular dynamics simulations, we study the slow dynamics of
supercooled liquids confined in a random matrix of immobile obstacles.
We study the dynamical crossover from glass-like to Lorentz-gas-like
behavior in terms of the density correlation function, the mean square
displacement, the nonlinear dynamic susceptibility, the non-Gaussian
parameter, and the fragility. 
Cooperative and spatially heterogeneous dynamics are suppressed as the
obstacle density increases, which lead to
the more Arrhenius-like behavior in
the temperature dependence of the relaxation time. 
Our findings are qualitatively consistent with the results of 
recent experimental and numerical studies for various classes of 
spatially heterogeneous systems. 
We also investigate the dependence of the dynamics of mobile particles
on the protocol to generate the random matrix.
A reentrant transition from the arrested phase to the liquid phase as the
mobile particle density {\it increases} is observed for a class of protocols. 
This reentrance is explained in terms of the distribution of the volume
of the voids 
that are available to the mobile particles. 
\end{abstract}

\submitto{\JPCM}
\maketitle

\section{Introduction}
\label{introduction}

The transport properties of fluids in a heterogeneous environment
are of great importance in physics, chemistry, and
engineering~\cite{Havlin2002Diffusion, AlbaSimionesco2006Effects}.
These systems include fluids confined in walls, standing
thin-film liquids, and fluids adsorbed in random media.
The effect of spatial confinement is especially important 
to the study of the glass transition of supercooled liquids. 
Recent experiments and computer simulations have revealed that 
the glass transition temperature, transport properties, and microscopic
dynamics sensitively change in the presence of spatial
confinement~\cite{Alcoutlabi2005Effects}. 
Moreover, it is expected that 
an understanding of these phenomena may lead to a deeper insight 
into the growing length scales of the cooperative motion of atoms,
which escorts the drastic slowing down of dynamics near the 
glass transition point~\cite{Scheidler2002Cooperative, Biroli2008Thermodynamic}.
A fluid in randomly distributed immobile obstacles is an ideal model 
to study the effects of the confinement on the glassy slow dynamics, 
in fact, a fluid in random media is interesting in its own right.
This system is introduced as a model system of a crowded environment, 
and the transport phenomena contained within has attracted a great deal of
attention in the
biophysics community~\cite{Saxton1994Anomalous,
Saxton1997Single, Ellis2003Join, Sung2008Lateral}. 
This system 
can also be seen as a generalization of the Lorentz gas
problem to the multi-particle case~\cite{Hofling2006Localization, Hofling2007Crossover,
Hofling2008Critical, Bauer2010The}.
Furthermore, this system can be regarded as a model of binary mixtures with a
disparate size ratio~\cite{Imhof1995Experimental, Dinsmore1995Phase,
Moreno2006Anomalous, Moreno2006Relaxation, Kikuchi2007Mobile, 
Horbach2002Dynamics, Voigtmann2006Slow, Mayer2009Multiple,
Voigtmann2009Double, Horbach2010Localization, Kurita2010Glass}. 
The immobile obstacles in the random media are interpreted as large
particles in the binary liquids because of the huge asymmetry in time
scales between the small and large particle components. 

In recent years, various molecular dynamics (MD) simulations have been performed to
study the slow dynamics of the fluids in random
media~\cite{Gallo2002Mode, Gallo2003Slow, Gallo2003Slow2, Kim2003Effects, Chang2004Diffusion,
Mittal2006Using, Sung2008The}.
Theoretically, the slow dynamics of mobile hard spheres in the presence of the
immobile hard spheres of the same size has been intensively investigated using
the replica method combined with the mode-coupling theory
(RMCT)~\cite{Krakoviack2005LiquidGlass, Krakoviack2005Liquidndashglass,
Krakoviack2007Modecoupling, Krakoviack2009Tagged}.
These studies have examined the dynamic phase transition from liquid to non-ergodic
arrested states.
Two notable results were predicted.
The first is that the slow dynamics can be characterized by
two types of dynamics: Type $A$ and Type $B$ dynamics~\cite{GotzeinHansen1991Liquids}.
When the mobile particle density, $\rho_{m}$, is large and
the immobile particle density, $\rho_{i}$, is small, the
system undergoes a conventional glass transition, in which the onset of
slow dynamics is signaled by the {\it discontinuous} emergence 
of two-step relaxation in the density correlation function.
This is referred to as Type $B$ transition.
As the immobile density $\rho_{i}$ increases, the glass transition
point of the mobile particles
decreases drastically.
At an even larger $\rho_{i}$, the dynamics 
qualitatively changes; one-step slow relaxation sets in at large
wavelengths, in which the tail of the relaxation curve continuously grows 
and progressively propagates toward the smaller wavelengths as
$\rho_{i}$ increases, until the dynamics of the mobile particle
freezes. 
This freezing is a localization transition due to blocking by
the percolating network of the immobile particles. 
This is called Type $A$ transition.
The crossover from Type $B$ to Type $A$ transition is 
reminiscent of the 
behaviors observed in many heterogeneous systems such as the glass-to-gel
crossover of attractive colloids~\cite{Zaccarelli2006Gel, Zaccarelli2007Colloidal}
and the glassy slow dynamics of binary mixture systems with
a disparate size ratio~\cite{Moreno2006Anomalous, Moreno2006Relaxation}.
The second prediction is the existence of a reentrant transition in
the small $\rho_m$ regime, in which the arrested mobile particles melt 
as the mobile particle density $\rho_m$ {\it increases} at a fixed $\rho_i$.

\begin{figure}[t]
\centering
\includegraphics[width=.9\textwidth]{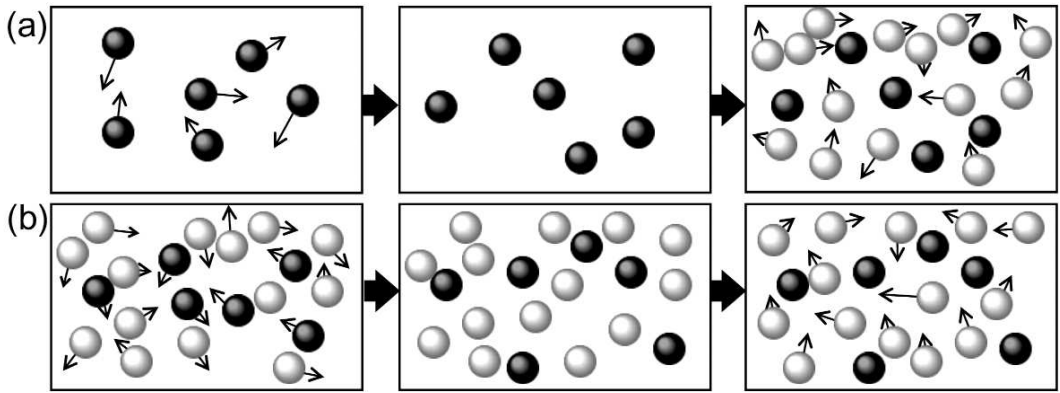}
\caption{Schematic illustrations of two protocols to generate the random matrix:
(a) Quenched-Annealed (QA) and (b) Equilibrated Mixture (EM) protocols.
}
\label{cartoon}
\end{figure}

To verify these theoretical predictions,
we~\cite{Kim2009Slow, Kim2010Molecular} and Kurzidim {\it et
al.}~\cite{Kurzidim2009Single, Kurzidim2010Impact} have independently
and numerically  
investigated fluids in immobile obstacles. 
All of these studies have confirmed that there is a crossover from Type $B$ to
Type $A$ transition as the immobile particle density increases. 
These studies also found no reentrant transition for the system that was
studied by RMCT.

We have also found that the dynamic arrest line (or more precisely
iso-relaxation-time line) sensitively depends on the protocol that is
used to
generate the configuration of the randomly distributed immobile
particles~\cite{Kim2009Slow, Kim2010Molecular}.
Two types of these protocols have been studied; the first is 
the Quenched-Annealed (QA) system, in which the immobile particles are
initially equilibrated before their positions are quenched (see
Fig.~\ref{cartoon}(a)).
Mobile particles are then inserted into the system and their dynamics
are monitored after equilibration.
This QA protocol offers a natural choice to model the experimental setups of
random media (such as porous materials) and has been adopted in the
theoretical analysis of
RMCT~\cite{Krakoviack2005LiquidGlass, Krakoviack2005Liquidndashglass,
Krakoviack2007Modecoupling, Krakoviack2009Tagged}
and in numerical studies by Kurzidim {\it et al.}~\cite{Kurzidim2009Single,
Kurzidim2010Impact}.
The second protocol studied is the Equilibrated Mixture (EM)
protocol, in which all of the particles are run in the simulation box and, after
their equilibration, the motions of a fraction of the particles are quenched
(see Fig.~\ref{cartoon}(b)).
The dynamics of the mobile component is monitored after waiting long
enough for the mobile particles to equilibrate in the presence of the
immobile particles. 
This protocol is appropriate as a model of the dynamics of the
fast (small) particle component in binary mixtures with a
disparate size ratio, in which the two time scales of each component are
well separated.
We found no reentrance for the system prepared with the QA protocol, but 
surprisingly, we did observe reentrance using the EM
protocol~\cite{Kim2009Slow, Kim2010Molecular}. 
It should be noted that this reentrance is distinct from that predicted by RMCT.
It was speculated that this reentrance can be attributed solely to the
configurations of the immobile particles prepared by the EM
protocol; 
the configuration of immobile particles is automatically ``optimized'' to
provide more pathways for the mobile particles than those prepared in the
absence of mobile particles (using the QA protocol). 

In the present paper, we investigate the dynamical properties of the
fluids in random obstacles (which were also examined in our previous
papers~\cite{Kim2009Slow, Kim2010Molecular}) in more
detail, focusing on various quantities that characterize the slow
dynamics near the glass and the localization transition. 
We evaluate the nonlinear dynamic susceptibility, the non-Gaussian parameter, and
the fragility across the entire parameter space of $(\rho_{i}, \rho_{m})$. 
We also quantify how the 
dynamics of mobile particles sensitively depends on
the protocol used to generate the random matrices.

This paper is organized as follows.
In Sec.~\ref{model}, 
we briefly review our model and simulation method.
In Sec.~\ref{result}, the numerical results are given;
we first describe how the dynamics changes from Type $B$ to Type $A$ by
calculating various dynamic quantities.
In the latter subsection of Sec.~\ref{result}, we discuss the sensitivity
of the dynamics of mobile particles to a geometry of random
configurations of the immobile particles by calculating the pore-size
distribution.
In Sec.~\ref{conclusion}, 
we summarize our results and provide our concluding remarks.

\section{Model and methods}
\label{model}

We perform MD simulations for two types of systems: a binary mixture
interacting with the soft-core potential and a monodisperse hard sphere. 
The binary mixture is used to explore the entire parameter space
of $(\rho_{i}, \rho_{m})$; bidispersity is required to avoid
crystallization at the small obstacle density limit $\rho_i\rightarrow 0$.
The monodisperse hard sphere is used to investigate the region at which
the crossover from Type $B$ to Type $A$ transition takes place. 
This is also the region where the reentrant transition was predicted by
RMCT. 
Monodispersity is of importance to explore the physical mechanism
near the crossover without the risk of it being obscured by the softness
of the potential or by the bidispersity of the system. 

The binary soft-core mixture consists of an equal number
of two types of particles with the total number $N = N_1 + N_2 = 500 + 500$.
They interact via the soft-core potential
\begin{equation}
v_{\alpha\beta}(r) = \epsilon\left(\frac{\sigma_{\alpha\beta}}{r}\right)^{12},
\end{equation}
where
$\sigma_{\alpha\beta} =(\sigma_\alpha + \sigma_\beta)/2$ and $\alpha,\beta \in \{1, 2\}$.
The size and mass ratio were $\sigma_2/\sigma_1 = 1.2$ and 
$m_2/ m_1= 2$, respectively.
The total number density was fixed at
$\rho=(N_1+N_2)/L^3=0.8\sigma_1^{-3}$, in which the
system length was $L=10.77\sigma_1$ under periodic boundary conditions (PBC).
The units of length, time, and temperature were considered to be $\sigma_1$,
$\sqrt{m_1 \sigma_1^2 / \epsilon}$, and $\epsilon / k_B$,
respectively.
For each simulation run,  
{$N_i$} particles were picked up from the $N$ particles randomly
and fixed their positions. $N_m=N-N_i$ particles were left mobile. 
In this model, we used the number densities defined by $\rho_i=N_i
\rho_{\eff}/N$ and
$\rho_m=N_m\rho_{\eff}/N$ as the system parameters, in which
$\rho_{\eff} =N(\epsilon/k_BT)^{1/4}{\sigma_{\rm eff}}^3/V$ is the
effective density of this model~\cite{Yamamoto1998Dynamics}.  
Here, $\sigma_{\eff}$ is the effective particle diameter defined by
${\sigma_{\rm eff}}^3 = \sum_{\alpha,
\beta=1,2}x_{\alpha}x_{\beta}{\sigma_{\alpha\beta}}^3$,
where $x_1 = N_1 / N = 1/2$ and $x_2 = N_2 / N = 1/2$ are the
mixture compositions.
The states that were investigated here were as follows:
$\rho_{\eff}=0.5$, $0.6$, $0.7$,
$0.8$, $0.9$, $1.0$, $1.1$, $1.15$, $1.2$, $1.3$, $1.4$, and $1.45$.
The corresponding temperatures were as follows: $T=21.61$, $10.42$,
$5.624$, $3.297$, $2.058$, $1.350$, $0.992$, $0.772$, $0.651$, $0.473$,
$0.352$, and $0.306$, respectively.
We controlled $\rho_i$ and $\rho_m$ by changing $N_i$ (or $N_m$) and
$\rho_{\eff}$ (or $T$). 
The number of mobile particles was chosen to be $N_m=10 \sim 900$.
The velocity Verlet algorithm was used to integrate Newton equations 
with time steps of $0.001\sim 0.005$.

The monodisperse hard sphere system includes $N$ identical hard spheres
with mass $m$ and diameter $\sigma$ in a cubic box of volume $V$ under PBC.
$\sigma$ and $\sqrt{{m\sigma^2}/{k_BT}}$ were used as the units of length
and time, respectively.
The temperature was fixed as $k_BT=1$. 
The standard event-driven algorithm was used for particle
collisions~\cite{Allen1989Computer}.
The number densities $\rho_i=N_i/V$ and $\rho_m=N_m/V$ were controlled
by changing $N_i$, $N_m$, and $V$. 
For both systems, we carefully checked the system size dependence 
and the sample dependence of the observables throughout the study.
Two types of protocols, the QA and EM protocols, were employed to generate
the random matrices.

\section{Numerical results}
\label{result}

\begin{figure}[t]
\centering
\includegraphics[width=.45\textwidth]{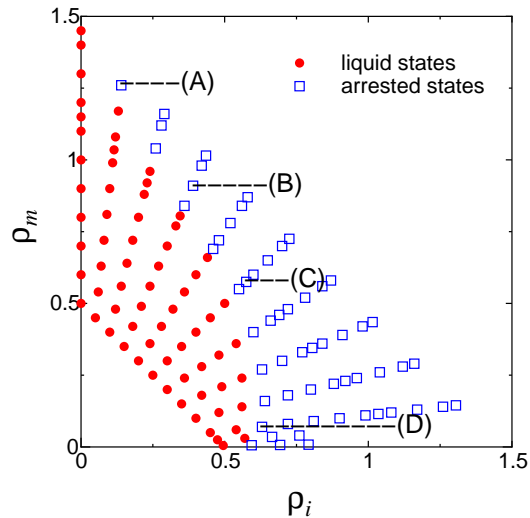}
\caption{
Dynamic phase diagram of the binary soft-core mixture generated by
the EM protocol.
$\rho_i$ is the immobile (obstacle) particle density, and 
$\rho_m$ is the mobile (fluid) particle density.
The arrested states are defined as the points beyond which the
$\alpha$-relaxation time $\tau_\alpha$ exceeds $10^3$.
}
\label{3D_soft_phase}
\end{figure}

\begin{figure}[t]
\centering
\includegraphics[width=.9\textwidth]{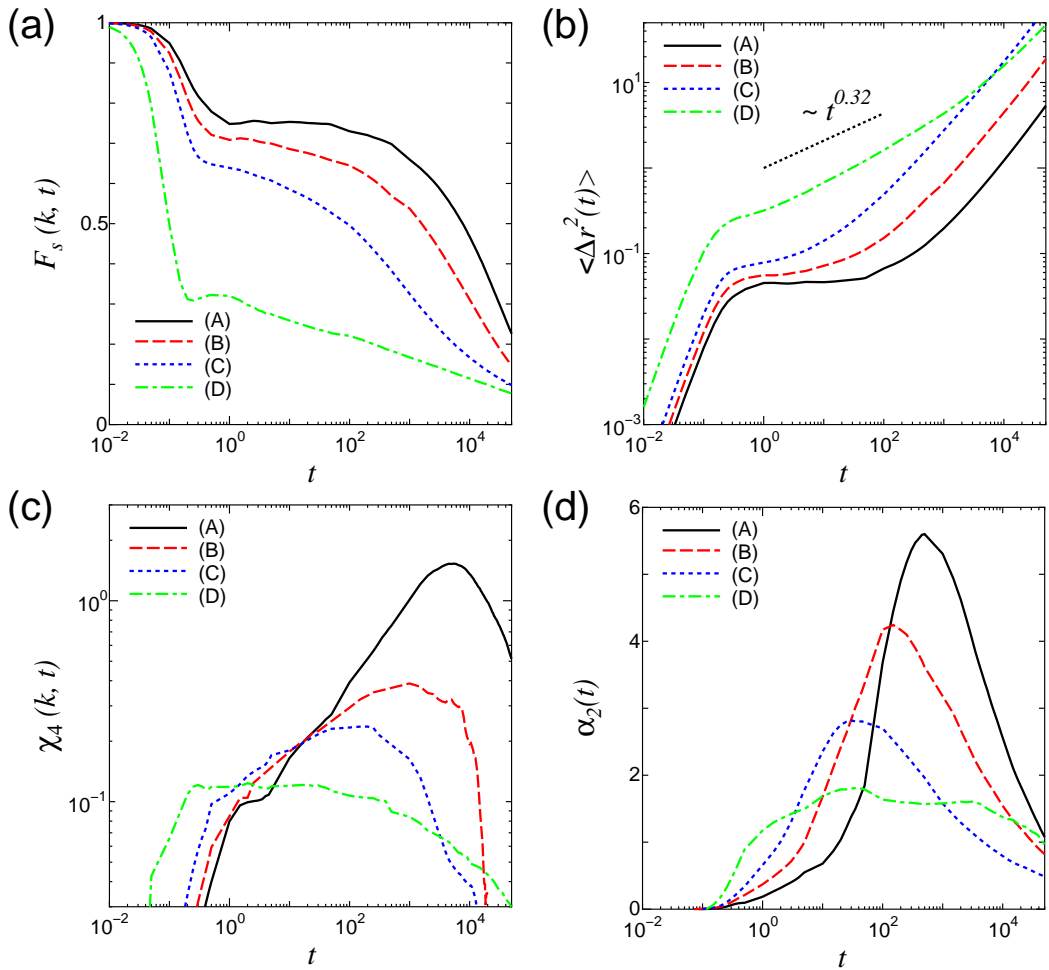}
\caption{
Various time-dependent quantities for the binary soft-core mixture
 generated by the EM protocol at the four state points (A) $(N_i,
 \rho_{\eff}) = (100, 1.45)$, 
(B) $(N_i, \rho_{\eff}) = (300, 1.3)$, 
(C) $(N_i, \rho_{\eff}) = (500, 1.15)$, and
(D) $(N_i, \rho_{\eff}) = (900, 0.7)$, as denoted in Fig.~\ref{3D_soft_phase}. 
(a) the self-part of the intermediate scattering function
$F_s(k, t)$ with $k=2\pi$,
(b) the	mean square displacement $\langle \Delta r^2(t)\rangle$,
(c) the nonlinear dynamic susceptibility $\chi_4(k, t)$ with $k=2\pi$,
and (d) the non-Gaussian parameter $\alpha_2(t)$.
}
\label{3D_soft}
\end{figure}

\subsection{Dynamic phase diagram}
\label{sec:PhaseDiagram}

We first determined the dynamic phase diagram for the whole $(\rho_{i},
\rho_{m})$-space by performing MD simulations of the EM systems of the
binary soft-core mixture.
In Fig.~\ref{3D_soft_phase}, the dynamic phase diagram is plotted  as a
function of $\rho_{i}$ and $\rho_{m}$. 
The dynamic arrest line is defined as the points at which the $\alpha$-relaxation time 
$\tau_\alpha$ reaches $10^3$.
We confirmed that varying
the criteria for $\tau_\alpha$
simply shifts the dynamic arrest line back and forth, but that its
qualitative behavior remains intact. 
$\tau_\alpha$ is determined by calculating the self-part of the intermediate
scattering function for mobile particles,
\begin{equation}
F_s(k,t)=\frac{1}{N_{m}}\left\langle
\sum_{j=1}^{N_{m}}\exp[i\vec{k}\cdot(\vec{r}_j(t)-\vec{r}_j(0))]\right\rangle,
\end{equation}
where $\vec{k}$ is the wave vector, $k=|\vec{k}|$, and 
$\vec{r}_j(t)$ is the position of the $j$-th particle.
We defined $\tau_\alpha$ by $F_s(k=2\pi, \tau_\alpha)=0.1$.

As $\rho_{i}$ increases, the dynamic arrest line (or the glass
transition points) drastically decreases.  
These features were well documented in the previous simulations
studies~\cite{Kim2003Effects, Chang2004Diffusion, Mittal2006Using,Sung2008The} 
and in RMCT~\cite{Krakoviack2005LiquidGlass, Krakoviack2005Liquidndashglass,
Krakoviack2007Modecoupling, Krakoviack2009Tagged}.
This tendency is sustained up to the region, beyond which a small 
reentrant pocket is observed. 
This reentrance will be
discussed in a later section.

\subsection{Intermediate scattering function}
\label{sec:FSKT}

The numerical results of conventional dynamical quantities displayed in 
Fig.~\ref{3D_soft} demonstrate how the immobile particle density
$\rho_i$ affects the relaxation processes of mobile particles.
%To investigate how the immobile particle density $\rho_i$ affects
%the relaxation processes of mobile particles, numerical results of
%conventional dynamical quantities are displayed in Fig.~\ref{3D_soft}.
In Fig.~\ref{3D_soft}(a), the time evolution of $F_s(k, t)$ with
$k=2\pi$ is plotted 
for the four state points (A)--(D) that are indicated in
Fig.~\ref{3D_soft_phase}, at which the $\alpha$-relaxation times are almost
the same.
The figure clearly indicates
that there are two types of distinct dynamics 
depending on $\rho_i$.

In the small $\rho_i$ (immobile particle density) regime $\rho_i \ll 0.5$,
$F_s(k,t)$ exhibits two-step relaxation with a well-developed plateau,
which is a hallmark of 
slow dynamics near the glass transition point.
We found that the shoulder of the plateau discontinuously appears as one approaches from
the fluid side to the arrested phase~\cite{Kim2009Slow, Kim2010Molecular}. 
This behavior is typical of slow dynamics near the glass transition point and
is referred to as Type $B$ transition in the MCT community.
However, as the mobile particle density $\rho_m$ decreases and 
$\rho_i$ increases,
the relaxation profile of $F_s(k, t)$ becomes quite different from that of
Type $B$ transition, {\it i.e.},
$F_s(k, t)$ shows a single step relaxation with a long tail (see
$F_s(k,t)$ at the state (D)).
It is also observed that the amplitude of the tail increases 
continuously as one crosses the arrested phase and that this 
increase incipiently starts from the lowest wavelength 
and propagates to the shorter scales as $\rho_i$
increases~\cite{Kim2009Slow, Kim2010Molecular}.    
This behavior is known as the hallmark of Type $A$ transition (or the localization
transition) as predicted by RMCT~\cite{Krakoviack2005LiquidGlass,
Krakoviack2005Liquidndashglass, Krakoviack2007Modecoupling,
Krakoviack2009Tagged} and demonstrated by simulations for various
spatially heterogeneous systems, such as binary mixtures of large and
small particles~\cite{Moreno2006Anomalous, Moreno2006Relaxation} and
colloidal gels~\cite{Zaccarelli2006Gel, Zaccarelli2007Colloidal}.

\subsection{Mean square displacement}
\label{sec:MSD}

The qualitative change from Type $B$ to Type $A$ dynamics is also
observed in the mean square displacement (MSD) for mobile particles,
\begin{equation}
\langle \Delta r^2(t) \rangle = \frac{1}{N_m}\left\langle
\sum_{j = 1}^{N_m} |\vec{r}_j(t)-\vec{r}_j(0)|^2
\right\rangle.
\end{equation}
The results are plotted in Fig.~\ref{3D_soft}(b).
It is known that in the Type $B$ dynamics, the MSD exhibits a plateau
at the $\beta$-relaxation time regime where the plateaus are observed
for $F_s(k,t)$, in which the tagged particle is trapped by surrounding particles.
However, 
at the small $\rho_{m}$ limit (see (D) of Fig.~\ref{3D_soft}(b)), 
anomalous subdiffusive behavior $\Delta {r}^2(t) \sim t^{\alpha}$ 
($\alpha < 1$) is observed, in which the system is almost Lorentz-gas-like.
On a short timeframe, the mobile particles at the state point (D) can explore
longer distance than
those at (A)--(C), because the mobile particle density is small. 
At a longer timeframe, however, the diffusion becomes very slow due to the 
developing network of immobile particles 
that hinders the ability of the mobile particles to explore the long distance;
this system
can be explained in terms of the percolation
theory~\cite{Stauffer1994Introduction}.
The subdiffusion exponent $\alpha \approx {0.3}$  is consistent
with  $0.32$ predicted for the Lorentz
gas~\cite{Hofling2006Localization,
Hofling2007Crossover,Hofling2008Critical, Bauer2010The}.   

\subsection{Nonlinear dynamic susceptibility}
\label{sec:DH}

We next investigate the nonlinear dynamic susceptibility or four-point 
correlation function for mobile particles, $\chi_4(k, t)$, 
which is a measure that is used to quantify the extent of the dynamic
heterogeneities~\cite{Glotzer2000Timedependent, Toninelli2005Dynamical,
Biroli2006Inhomogeneous}.
$\chi_4(k, t)$ is defined as the variance of the 
fluctuations of the self part of the intermediate scattering function by 
\begin{equation}
\chi_4(k, t) = N_{m} [ \langle \hat F_s^2(k, t) \rangle- \langle
\hat{F_s}(k, t)\rangle ^2 ].
\end{equation}
Here $F_s(k, t)=\langle\hat F_s(k, t)\rangle$ and 
\begin{equation}
\hat F_s(k, t)\equiv
 \frac{1}{N_{m}}\sum_{i=1}^{N_{m}}\frac{\sin(k|\Delta\vec{r}_i(t)|)}{k|\Delta\vec{r}_i(t)|}.
\label{eq:DH1}
\end{equation} 
In the literatures~\cite{Toninelli2005Dynamical}, 
$N_{m}^{-1}\sum_{i=1}^{N_{m}}\cos(\vec{k}\cdot\Delta\vec{r}_i(t))$ is
conventionally used as the definition of $\hat F_s(k,t)$. 
Under this definition, $\chi_4(k, t)$ decays to a constant $1/2$ at
$t\rightarrow \infty$.
However, as we demonstrate here,
the peak of $\chi_4(k,t)$ for the Type $A$ regime grows more mildly than
it does for the Type $B$ regime. 
To demonstrate this suppression of the peak and thus the dynamic heterogeneities 
in the Type $A$ regime without the results
being obscured by a constant plateau at a large $t$,  
we have adopted an alternative definition of $\hat F_s(k, t)$ by taking
the average over the angular components of the wave vector $\vec{k}$,
which leads to eq.(\ref{eq:DH1}).
Note that both definitions of $\hat F_s(k,t)$ lead to an identical 
averaged value $F_s(k,t)=\langle \hat F_s(k,t)\rangle$ due the isotropic
nature of the system, but that the new definition removes the unwanted
constant for $\chi_4(k, t)$ at $t \to\infty$.
In Fig.~\ref{3D_soft}(c), the time evolutions of $\chi_4(k, t)$ are plotted for
the four state points (A)--(D).
In the Type $B$ regime, $\chi_4(k, t)$ exhibits behavior that is typical
for bulk glass, {\it i.e.}, a pronounced peak at the $\alpha$-relaxation 
time, whose height grows rapidly as the density increases and is 
preceded by algebraic growth in the $\beta$-relaxation regime. 
In the Type $A$ regime at state (D), however,
$\chi_4(k, t)$ neither grows nor shows a strong peak, even after a long
period of time. 
This results implies that dynamic heterogeneities play a minor role in
the slow dynamics of this regime.
Similar behavior of $\chi_4(k, t)$ has been reported for colloidal gels
whose slow dynamics are caused by geometrical
constraints~\cite{Abete2007Static, Fierro2008Dynamical}.

\subsection{Non-Gaussian parameter}
\label{sec:NGP}

The non-Gaussian parameter (NGP) is another typical quantity that is
suitable to monitor the effect of the heterogeneities inherent in the system.  
We calculated the NGP $\alpha_2(t)$ defined by
\begin{equation}
\alpha_2(t)=\frac{3\langle\Delta r^4(t)\rangle}{5\langle\Delta
r^2(t)\rangle^2}-1,
\end{equation}
where $\langle \Delta r^4(t)\rangle = (1/N_m)\langle \sum_{j=1}^{N_m}
 |\vec{r}_j(t)-\vec{r}_j(0)|^4 \rangle$.
$\alpha_2(t)$ reveals how the distribution of the single-particle
displacement, $|\Delta \vec{r}_j(t)|$, at time $t$ deviates from the Gaussian
distribution~\cite{Hansen2006Theory}.
The profiles of $\alpha_2(t)$ are indicated in
Fig.~\ref{3D_soft}(d).

It is observed that $\alpha_2(t)$ develops and exhibits the pronounced
peak 
before the $\alpha$-relaxation time in the Type $B$ regime.
It is known that an increase in the maximum $\alpha_2(t)$ is synchronized with the
growing {\it dynamic} heterogeneities near the glass transition
point~\cite{Kob1997Dynamical}. 
As the immobile density $\rho_i$ increases, the height of the peak decreases. 
At the largest $\rho_i$, the point (D), it is hard to see the peak. 
This trend is qualitatively similar to that of the nonlinear dynamic
susceptibility, $\chi_4(k,t)$. 
Recently, Flenner and Szamel have proposed a new non-Gaussian parameter
(NNGP)~\cite{Flenner2005RelaxationBD}. 
They argued that the conventional NGP is 
more susceptible to particles moving faster than
those moving more slowly, 
whereas NNGP is more susceptible to slowly moving particles. 
It would be beneficial to compute and compare the NGP, NNGP, and $\chi_4(k,t)$ 
in order to corroborate the role of 
the dynamic and static heterogeneities in the confined systems.

\subsection{Fragility}
\label{sec:fragility}

\begin{figure}[t]
\centering
\includegraphics[width=.9\textwidth]{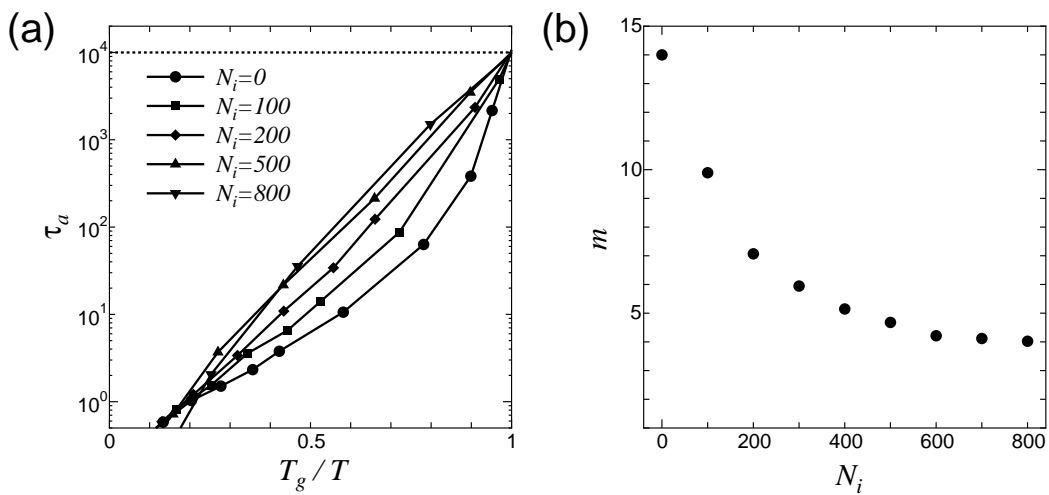}
\caption{
(a) The $\alpha$-relaxation time, $\tau_\alpha$, as a function of the
 inverse temperature $T_{g}/T$, 
for several $N_i$'s, $N_i=0$, $100$, $200$, $500$ and $800$.
$T_g$ is defined as the temperature at which $\tau_\alpha=10^4$. 
(b) $N_i$-dependence of the fragility index $m$.
}
\label{fragility}
\end{figure}

The ``fragility'' is a concept 
to quantify the deviation of the temperature dependence of the
viscosity, diffusion coefficient, and the relaxation time from the
Arrhenius behavior~\cite{Angell1991Relaxation}.
The fragility index $m$ is commonly used as a measure of the fragility
and is defined by the steepness of the increase
of $\tau_\alpha$ upon decreasing the temperature;
\begin{equation}
m = \frac{\partial \log_{10} \tau_\alpha}{\partial  (T_g/T)}\biggm|_{T=T_g},
\end{equation}
where $T_g$ is the glass transition point.
$m$ depends on the material of the glass formers.
Generally, fragile liquids with large $m$'s tend to exhibit more 
pronounced and more temperature-sensitive dynamic heterogeneities than
do more Arrhenius-like fluids with smaller $m$'s (or stronger
liquids).
As was demonstrated in the previous sections, the dynamic
heterogeneities are suppressed as the immobile particle density
$\rho_i$ increases.
Therefore, it is natural to expect that the system concomitantly becomes
stronger (more Arrhenius-like). 
We examined the temperature (or the effective density $\rho_{\eff}$)
dependence of $\tau_\alpha$'s for various $N_i$'s (the number of the
immobile particles) for a binary soft-core mixture.
Note that we have used $\rho_{\eff}$ to control $T$. 
The change of $\rho_{\eff}$ changes $\rho_i$ (and also $\rho_m$) slightly,
and this shift may make it difficult to quantify 
the effects of the fixed number of obstacles on the fragility.
We believe, however, that this effect is negligible for the range of the
temperatures that we have explored.

The ``Angell-plot'', the $\log_{10}\tau_{\alpha}$-vs-$1/T$ plot, 
of our system is shown in Fig.~\ref{fragility}(a), in which the
temperature $T$ is scaled by the ``glass transition temperature'' $T_g$.
$T_g$ has been defined by the point at which $\tau_\alpha$ reaches
$10^4$ that is slightly longer than the criteria used to draw
Fig.~\ref{3D_soft_phase}.
The dependence of the fragility index $m$ on the number of immobile
particles $N_i$ is plotted in Fig.~\ref{fragility}(b).
We observe that $m$ is remarkably sensitive to the
density of immobile particles. 
The fragility index is $m \approx 14$ for bulk glass but it
decreases to $m \approx 4$
at the largest density of immobile particles.
Our observation differs from the one reported for a binary
mixture system with large size ratios ($\leq 3$) by Kurita {\it et
al.}~\cite{Kurita2010Glass}.
They indicated that the fragility changed non-monotonically, though
mildly, by changing the size ratio of small and large particles and their
densities.  
It would be interesting to study how this trend changes as the size
ratio of the two components increases, resulting in
the time scales for each
component becoming decoupled. 

We remark that the fragility, which behaves in a manner that is
qualitatively similarly to
ours, has been experimentally obtained in polymeric systems confined in
porous media recently~\cite{Schonhals2007Segmental}, in which the crossover
from a non-Arrhenius to Arrhenius temperature-dependence of the
relaxation time was observed as the pore size became smaller. 
It was speculated that the decrease of the fragility as the effect of the
confinement is enhanced is universal and should be observed for other
types of confined systems such as those with the solid-liquid or air-liquid
interfaces~\cite{Inoue2009Glass}.

Finally, it should be noted that even the largest values of $m$ reported
here are still very small 
compared with conventional molecular systems~\cite{Bohmer1993Nonexponential}. 
This discrepancy occurs because the glass transition temperature $T_g$
defined above is
far higher than those observed for real glasses, which is due to 
the limited time windows that the simulations can access. 
It is noteworthy that our simulation results still exhibited qualitatively
similar behavior for $m$ as the experimental results, despite the
large time-scale differences between them.

\subsection{Reentrant transition}

\begin{figure}[t]
\centering
\includegraphics[width=.9\textwidth]{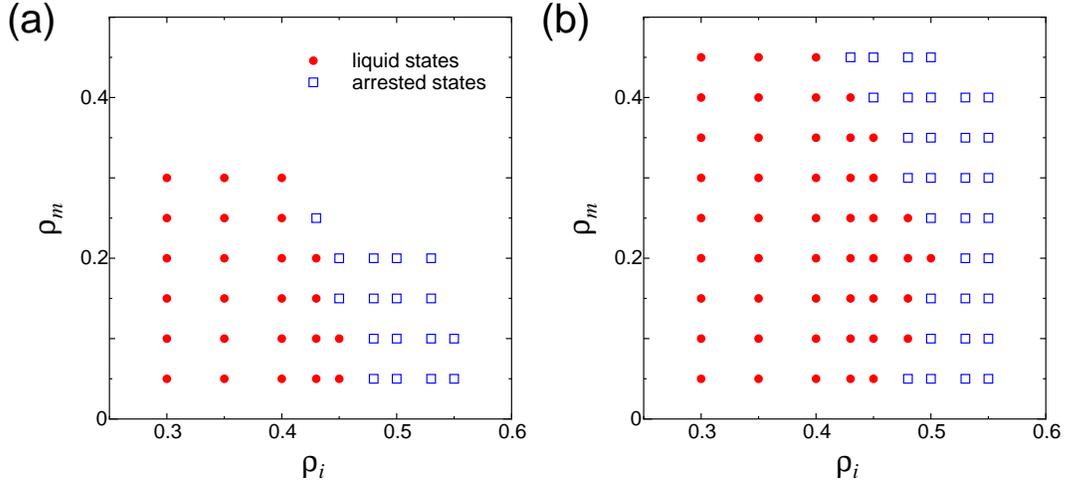}
\caption{
Dynamic phase diagram of the monodisperse hard sphere system with 
a random matrix generated by the two different protocols: (a) QA and (b) EM.
}
\label{3d_hard_phase}
\end{figure}

\begin{figure}[t]
\centering
\includegraphics[width=.9\textwidth]{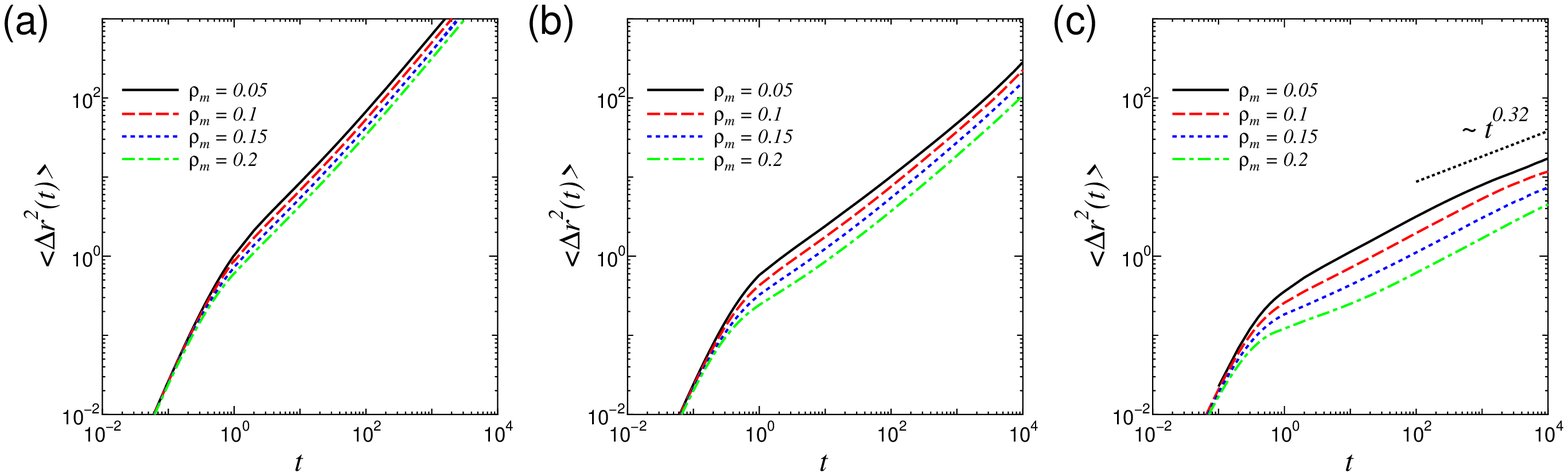}
\caption{
The $\rho_m$ dependence of the mean square displacement, $\langle \Delta r^2
(t)  \rangle$, at the immobile particle density
(a) $\rho_i=0.3$, (b) $0.43$, and (c) $0.5$ of the QA systems
}
\label{3d_hard_QA_msd}
\end{figure}

\begin{figure}[t]
\centering
\includegraphics[width=.9\textwidth]{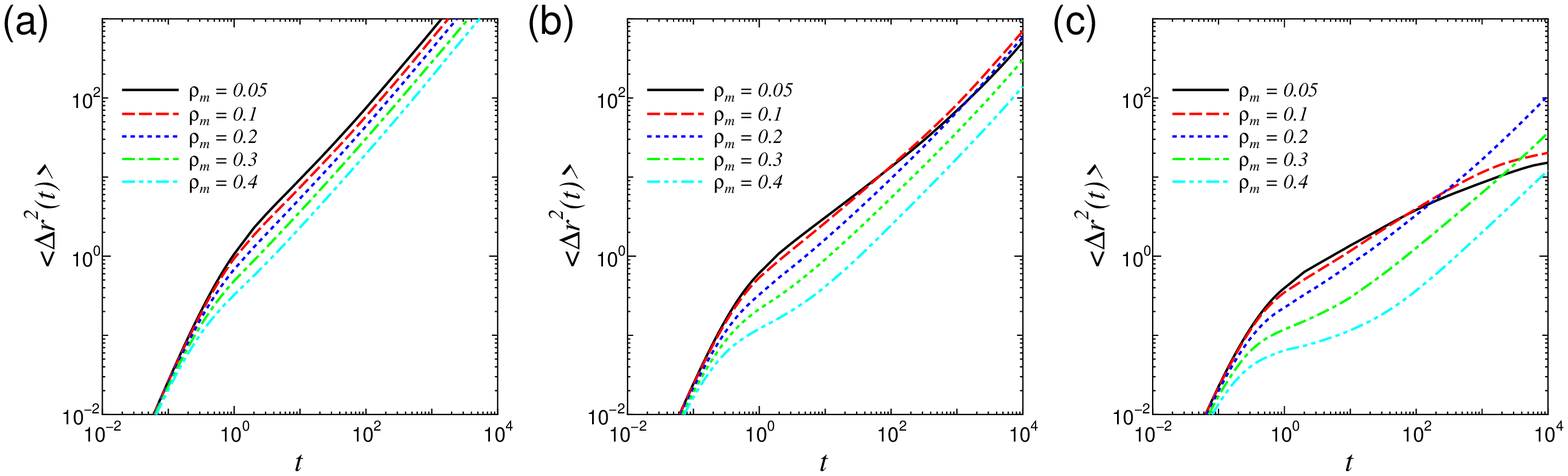}
\caption{
The $\rho_m$ dependence of the mean square displacement, $\langle \Delta
 r^2 (t) \rangle$, at the immobile particle density (a) $\rho_i=0.3$,(b)
 $0.43$, and (c) $0.5$ of the EM systems.}
\label{3d_hard_EM_msd}
\end{figure}

In this subsection, we examine the effect of the configurations of
immobile particles on dynamics of the mobile particles by comparing results obtained with
the EM protocol to those obtained with the QA protocol in the monodisperse hard sphere system.
We used
the one-component hard spheres to study the
mechanism behind the configuration-dependence of dynamics near the Type
$B$-Type $A$ crossover without the results being obscured by the softness of the
potential or by the bidispersity of the system. 
We calculated the MSD of the mobile particles and determined the
dynamic phase diagrams for both the QA and EM systems.
The results are plotted in Fig.~\ref{3d_hard_phase}.
Here, the dynamic arrest line is determined as the points at
which the MSD reaches $10^2$ in the simulation time $t=10^4$.
Figs.~\ref{3d_hard_QA_msd} and \ref{3d_hard_EM_msd}
show the dependence of the MSD on $\rho_m$ 
at several $\rho_i$'s for both the QA and EM systems.

As indicated by Fig.~\ref{3d_hard_phase}(a), {\it no} reentrant transition
is observed for the QA system.
The dynamic arrest line monotonically decreases as $\rho_i$ increases,
which is compatible with the 
recent numerical simulations for the related QA
systems~\cite{Kurzidim2009Single, Kurzidim2010Impact}.
Indeed, Fig.~\ref{3d_hard_QA_msd} indicates that the slope of the MSD
monotonically decreases as $\rho_m$ increases at a fixed $\rho_i$.
On the one hand, 
this result is hardly surprising because the slowing down of the
mobile particle dynamics is mainly due to the geometrical confinement by 
the immobile particles.
On the other hand, the EM system shows the reentrant pocket at a finite
$\rho_m$, which is
clearly seen in Fig~\ref{3d_hard_phase}(b).
The similar reentrance pocket has been observed in the binary soft-core mixture
(see Fig.~\ref{3D_soft_phase}).
The dynamics of the mobile particles are {\it
accelerated} in spite of the increase in $\rho_m$.
This reentrance can be  clearly seen in Fig.~\ref{3d_hard_EM_msd}(b) and
(c); by increasing
$\rho_m$ at the fixed large $\rho_i$, the slope of the MSD at long times
first {\it increases} and then
gradually decreases.

\begin{figure}[t]
\centering
\includegraphics[width=.45\textwidth]{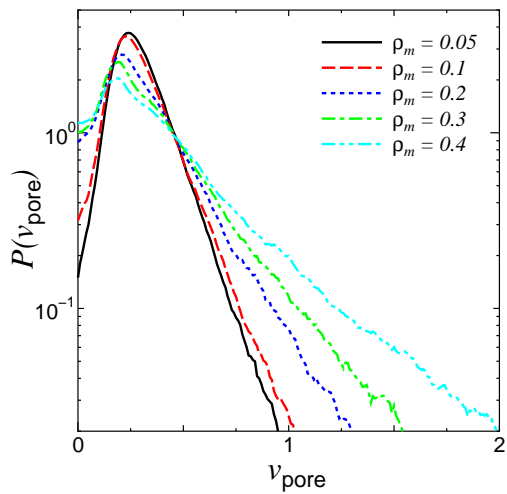}
\caption{
The pore-size distribution $P(v_{\rm pore})$ for various mobile particle
densities $\rho_m$ at the fixed immobile particle density $\rho_i=0.5$
of the EM systems.
}
\label{3d_hard_volume}
\end{figure}

In our previous study~\cite{Kim2009Slow, Kim2010Molecular}, we have
speculated that the origin
of this reentrance is due to the change of the equilibrium structure of
the immobile particles in the presence of the mobile particles, which
are equilibrated together when the random matrix is generated.
To verify this speculation, 
we investigated the distribution of the pore size (or the free volume
available for the mobile particles) generated by the
immobile particles. 
The pore-size distribution is determined as follows.
Using a three dimensional Delaunay triangulation algorithm, 
the total space of the system is divided into non-overlapping
tetrahedrons.
The vertices consist of the positions of the immobile particles.
The volume distribution of the tetrahedrons, $P(v_{\rm pore})$, for EM
systems is computed.
If the volume of the tetrahedron is much larger than that of the
particle, $v_{\rm pore} > \pi \sigma^3 / 6 \simeq 0.52\sigma^3$, 
the mobile particle can access the pore.
The available pore-sizes for the mobile particles are not identical to the 
available pathways that are available to them, but their distribution
function still
provides reliable information on the dynamics of the mobile
particles in geometrical confinement.
As observed in Fig.~\ref{3d_hard_volume},
the height of the tail of $P(v_{\rm pore})$ at $v_{\rm pore}\geq
0.52\sigma^3$ {\it increases} as $\rho_m$ {\it increases} at a fixed
value of $\rho_i=0.5$ monotonically.
This result indicates that the free volumes available for the mobile particles
increases, which leads to the reentrant behavior of the MSD that is
observed in Fig.~\ref{3d_hard_EM_msd}(c).
Note that the tails monotonically increase
(at least up to $\rho_m=0.4$),
but the dynamics slow down again
around $\rho_m\approx 0.2$, as indicated by Fig.~\ref{3d_hard_phase}(b).
This result occurs because the glassy dynamics of the mobile particles
sets in while the localization
effect (due to the geometrical confinements) becomes smaller.

These results quantify our speculation that the immobile particles
adjust themselves during an equilibration run to prepare for 
the presence of the mobile particles
so that the free volumes for both components are entropically maximized
and leave more available geometrical spaces (and pathways) for the
mobile particles, 
delaying the percolation transition to larger values and thus leading to faster
dynamics of the mobile particles.
The sensitivity of the percolation point to the protocols used to generate
the matrix configuration has been previously studied in several
contexts and our results are consistent with
these results~\cite{Chang2004Diffusion,Mittal2006Using,Sung2008The}. 

We also speculate that counterintuitive effects similar to the reentrance
discussed above are prevalent in systems in which
the disparate time scales are entangled. 
A binary mixture consisting of large and small 
particles studied by MD simulations may be a
good example. 
Voigtmann {\it et al.} have numerically studied the dynamics of such
a binary mixture and have shown that the diffusion of small particles 
becomes slower when the interactions between small particles are 
turned off~\cite{Voigtmann2009Double, Horbach2010Localization}.
One may speculate that this observation is relevant to
our finding of the reentrant pocket for the EM system.
The turn-off of the interactions makes small particles ``invisible'' to each
other and makes the large particles behave as if there are fewer small
particles around them, which makes the configuration of the large particles
become more QA-like than those with full interactions.
This effect might be difficult to observe via standard static quantities
(such as the static structure factors) 
but may be detected easily via the pore-size distribution function.
An accurate theoretical method to evaluate the subtle differences in
the static structure factors would be desirable to investigate how the
protocol dependence changes the dynamic phase diagram using
RMCT~\cite{Krakoviack2010Statistical}.

\section{Conclusions}
\label{conclusion}

In this paper, MD simulations have been performed to examine
the dynamical properties near the arrest points of simple fluids
confined in random media.
We calculated various quantities for the whole range of the mobile and
immobile densities, including the 
intermediate scattering function, the mean square displacement, the
nonlinear dynamic
susceptibility, the non-Gaussian parameter, and the fragility.
We found that all of these quantities exhibited qualitative changes as
the density of the mobile/immobile particles was varied. 
At the limit of small immobile particle density, 
all of the observed quantities indicated the strong signs of 
dynamic heterogeneities near the glass transition point, such as the enhanced
nonlinear dynamic susceptibility, the increased peak heights of the
non-Gaussian parameter,
and the fragile behavior of the relaxation time in its temperature
dependence. 
At the opposite limit, in which a small number of mobile particles
diffuse amid abundant obstacle particles, the signs of the dynamic
heterogeneities were all suppressed. 
The nonlinear dynamic susceptibility and the non-Gaussian parameter
exhibited no peak,
and the temperature dependence of the relaxation time was almost Arrhenius.
To understand  the underling physics behind the reentrant
transition near the Type $B$ to Type $A$ crossover, we have carefully quantified 
how the statistical properties of configuration of
the random matrix can be altered by the different protocols used to generate them
by calculating the pore-size distribution generated by the immobile particles.
Throughout this paper, we refer the change from Type $B$ to Type $A$
dynamics as the crossover. 
According to RMCT~\cite{Krakoviack2005LiquidGlass,Krakoviack2005Liquidndashglass,
Krakoviack2007Modecoupling,Krakoviack2009Tagged}, this change should be
associated with a higher order MCT transition. 
However, this transition is too subtle to be clearly observed at the
resolution of the current simulations.

\section*{Acknowledgments}
This work was partially supported by Grants-in-Aid for Scientific
Research: Young Scientists (B) (Grant No.~21740317),
Scientific Research (B) (Grand No.~22350013), Scientific Research (C)
(Grand No.~21540416) and Priority Area
``Soft Matter Physics.''
This work was also supported by the Center for the Promotion of
Integrated Sciences (CPIS) of Sokendai and 
the Next Generation Super Computing Project, Nanoscience Program, MEXT, Japan.
The numerical calculations were performed at Research Center for Computational Science,
Okazaki Research Facilities, National Institutes of Natural Sciences, Japan.

\section*{References}
%\bibliography{/home/kin/papers/kkims}
\providecommand{\newblock}{}

\end{document}